\documentclass[twocolumn,showpacs,preprintnumbers,amsmath,amssymb,10pt,a4paper]{revtex4}

\usepackage{geometry}
\usepackage{graphicx}
\usepackage{dcolumn}
\usepackage{bm}
\usepackage[english]{babel}

\newcommand{\be}{\begin{equation}}
\newcommand{\ee}{\end{equation}}
\newcommand{\BE}{\begin{eqnarray}}
\newcommand{\EE}{\end{eqnarray}}

\begin{document}

\preprint{}
\title{WKB versus generalized van Kampen system-size expansion: the stochastic logistic equation}
\author{Claudia Cianci}
\email{ccianci@staffmail.ed.ac.uk}
\affiliation{School of Biological Sciences, University of Edinburgh, Mayfield Road, Edinburgh EH9 3JR, UK}
\author{Duccio Fanelli}
\email{duccio.fanelli@unifi.it}
\affiliation{Dipartimento di Fisica e Astronomia, Universit\`{a} degli Studi di Firenze and INFN, via Sansone 1, IT 50019 Sesto Fiorentino, Italy}
\author{Alan~J. McKane}
\email{alan.mckane@manchester.ac.uk}
\affiliation{Theoretical Physics Division, School of Physics and Astronomy, 
The University of Manchester, Manchester M13 9PL, UK}

\begin{abstract}
Stochastic fluctuations are central to the understanding of extinction dynamics. In the context of population models they allow for the description of the transition from the vicinity of a non-trivial fixed point of the deterministic dynamics to a trivial fixed point, where the population has become extinct. To characterize analytically the fluctuations of a given stochastic population model, one can operate within the so-called linear-noise approximation. Here the fluctuations are taken to be Gaussian, and the phenomenon of extinction is so rare as to be negligible, for all practical purposes. When the size of the population becomes small, non-Gaussian fluctuations are instead found. Two analytical schemes are in principle available to determine the nature of the distribution of associated fluctuations: the generalized van Kampen system-size expansion beyond the conventional order of approximation and a WKB-like method. Here we investigate the accuracy of these two different approximation schemes, with reference to a simple stochastic process that converges to the logistic equation in the deterministic limit.
\end{abstract}


\maketitle

\section{Introduction}
\label{intro}
The huge increase recently seen in the construction of quantitative models in the biological sciences, especially ecology, has prompted a renewed interest in stochastic effects in these systems~\cite{MckaneEcologyTrends}. In computer simulations processes such as birth, death, predation, and many others, are quite naturally treated as random. In the mathematical versions of these individual-based models, the stochastic effects are present; they only vanish in the limit of infinitely many individuals, when deterministic equations are obtained. However, in some important situations, the stochastic effects can be so large as to invalidate the conclusions based on a purely deterministic analysis~\cite{McKanePRL,dipatti,McKane,De_Anna,McKaneBiancalaniRoger,woolley}. Hence the interest in stochastic dynamics, and the mathematical techniques which can be used to elucidate them.

The deterministic limit usually yields a set of coupled ordinary differential equations in time. These can be analysed using the well-developed ideas from the theory of dynamical systems involving concepts such as trajectories in phase space and attractors of various kinds. Stochastic systems have their own, different, set of concepts such as stochastic fluctuations about the deterministic trajectories, rare events which lead to transitions from one attractor of the deterministic dynamics to another, stationary probability distributions, and so on. It is the tools that can be used to calculate some of these quantities for a wide class of models that will of interest to us in this paper.

We will focus on extinction dynamics, that is, transitions from the vicinity of a non-trivial fixed point of the deterministic dynamics, to a trivial fixed point where all the individuals have become extinct. If the parameters of the models are chosen so that the two fixed points are not too close, then extinctions will be rare and the state of the system will continue to fluctuate about the non-trivial fixed point for a long period of time. In this situation the van Kampen system-size expansion~\cite{Vankampen} has proved to be a powerful tool. Essentially the method gives the deterministic equations to leading order in a expansion in (inverse) system-size, and linear stochastic corrections to this result at next-to-leading order. This linear approximation, sometimes called the linear-noise approximation, corresponds to Gaussian fluctuations, and is an excellent approximation if extinctions are so rare as to be negligible. 

In principle, extinction events can be incorporated into the system-size expansion by going beyond the Gaussian approximation, and including non-linear terms in the stochastic differential equation giving the approximate stochastic dynamics. This should give rise to a tail on one side of the probability distribution function which will characterise the extinction process. These higher-order calculations have only been carried out recently, and then only by a few authors~\cite{grima,grima2011,thomas,cianci_voter,cianci_epjst}. In addition if there are only a few individuals in the system, extinction effects will be very important, and one may have to go to quite high-order in the expansion to get an accurate form for what will be a very non-Gaussian distribution.

By contrast, the standard technique to look at rare events, found going from one metastable state to another, is to use a WKB-like approximation. This goes under many names: large deviation theory~\cite{touchette}, the instanton method~\cite{zinn-justin}, Freidlin-Wentzell theory \cite{FreidlinWentzell}, amongst others. It consists of postulating that the dominant contribution to the probability distribution is exponentially small in $N$, the system-size, that is, is of the order of $e^{-NS}$. Here $S$ is a function of variables describing the system in the deterministic limit which turns out to satisfy a Hamilton-Jacobi equation. The corresponding Hamiltonian, known as the Freidlin–Wentzell Hamiltonian, can be used to describe the extinction trajectories, even though they are stochastic in nature. 

The aim of this paper is to explore the connection between the van Kampen system-size expansion at higher-order and the WKB method. They have a very different basis, and to the best of our knowledge, their predictive power in regimes where extinctions are important have not been compared. We will carry out the explicit assessment of their range of validity and comparison with numerical simulations on a specific stochastic system with one degree of freedom to minimise numerical errors. In the case of the WKB method, most of the steps can be performed analytically, which also makes the interpretation of the results more straightforward. A simple dynamical system with a stable non-trivial fixed point and an unstable trivial fixed point is the logistic equation $\dot{\phi}=r\phi\left[ 1 - (\phi/k) \right]$, and the stochastic model we will choose to study will have this equation as its deterministic limit. 

\section{Model}
\label{model}
In this section we will introduce the individual based model we will analyse and write down the master equation, which governs its stochastic dynamics. The two techniques we are comparing can be viewed as different approximations to these dynamics. We will describe them in turn and then compare them to numerical simulations of the original individual based model in Section~\ref{simulations}.

Following the discussion in the Introduction, probably the simplest model which contains the features which we wish to explore is a system containing identical individuals, which we will denote by $A$. We assume that there are $n$ such individuals, and since the size of the system is taken to be characterised by an integer $N$, we suppose that there are $(N-n)$ nulls denoted by $E$. These are vacancies, which in a spatial version of the model would denote spaces which could potentially be colonised by a individual. If the only processes are (asexual) birth, competition and death, then we may define the well-mixed model through the reactions
\begin{equation}
A + E \xrightarrow{b} A + A, \ \  A + A \xrightarrow{c} A + E, \ \ 
A\xrightarrow{d}E. 
\label{reactions}
\end{equation}
The last equation, for instance, indicates that an individual of type $A$ dies at a rate $d$ to give a vacancy, $E$. Simple combinatorics then gives the rate at which the number of individuals increases from $n$ to $(n+1)$ to be given by $b\,(n/N)\,(N-n)/N$. A more accurate statement of these rates would replace one of the $N$ factors in the denominator by $(N-1)$, but since we wish to keep the analysis as simple as possible we will not do this. If we scale the time by a factor of $N$, then the transition rate from state $n$ to state $(n+1)$ may be written as:
\begin{equation}
T(n+1|n)=bn\left( 1 - \frac{n}{N} \right).
\label{n_to_n+1}
\end{equation}
Similarly, the transition rate from state $n$ to state $(n-1)$ is
\begin{equation}
T(n-1|n) = n \left( d + c \frac{n}{N} \right),
\label{n_to_n-1}
\end{equation}
where once again factors of $N(N-1)$ and $n(n-1)$ have been replaced by $N^2$ and $n^2$ respectively. Since the transition rates $T(n+1|n)$ and $T(n-1|n)$ define the model, this choice simply corresponds to a slight variant of the standard model, which can be justified on grounds of simplicity.

The master equation is an equation for the rate of change with time of the probability of finding $n$ individuals in the system at time $t$, denoted by $P(n,t)$. Since this is simply the rate of transitions into the state $n$ minus the rate of transitions out of state $n$, it reads~\cite{Vankampen}
\begin{eqnarray}
\frac{dP(n,t)}{dt} &=& T(n|n+1)P(n+1,t) \nonumber \\
&+& T(n|n-1)P(n-1,t) \\
&-& T(n-1|n)P(n,t) - T(n+1|n)P(n,t) \nonumber .
\label{master}
\end{eqnarray}
This equation cannot be solved exactly, so we need to resort to either numerical methods or approximation techniques. It is frequently simpler to simulate~\cite{gillespie} the processes given in Eq.~(\ref{reactions}), rather than numerically solve the master equation, and the results we give in Section~\ref{simulations} to assess the accuracy of the approximation techniques are found in this way. We now briefly outline the two approximation methods that we use in this paper.

\subsection{The van Kampen system-size expansion}
\label{vK_method}
The first, the van Kampen system-size expansion, has as the leading-order approximation the deterministic differential equation found by taking the limit $N \to \infty$. In the case of the model just described, this is the logistic equation given in the Introduction. However this equation emerges as the leading order approximation to the model defined by Eqs.~(\ref{n_to_n+1}) and (\ref{n_to_n-1}), and does not have to be postulated independently. The next-to-leading order gives a linear stochastic differential equation, which describes Gaussian fluctuations about the deterministic result. If the intention is to simply study the stochastic dynamics of the model well away from the boundaries, it is usually sufficient to work to this order. However it is possible to go higher orders to obtain non-Gaussian corrections to the probability distribution function (pdf). One of the main aims of this paper is to argue that these higher-order corrections enable reliable estimates for the pdf to be obtained very close to the boundaries. 

To apply the van Kampen expansion we first write down the master equation (\ref{master}) in terms of step-operators ${\cal E}^{\pm}$ defined by ${\cal E}^{\pm}f(n)=f(n\pm1)$, where $f$ is an arbitrary function~\cite{Vankampen}, 
\begin{eqnarray}
& & \frac{dP(n,t)}{dt} = \left( {\cal E}^{+} - 1 \right) 
\left[T(n-1|n)P(n,t) \right] \nonumber \\
& & + \left( {\cal E}^{-} - 1 \right) 
\left[T(n+1|n)P(n,t) \right].
\label{var_master}
\end{eqnarray}
The ansatz which forms the basis of the method is to write~\cite{Vankampen}
\begin{equation}
\frac{n}{N} = \phi(t) + \frac{\xi}{\sqrt{N}}.
\label{ansatz}
\end{equation}
Here $\phi(t)$ is the solution of the deterministic equation valid in the limit $N \to \infty$ and $\xi$ is the (continuous) stochastic variable which gives the deviation of the stochastic trajectory from this deterministic value. The pdf when written in terms of $\xi$ is denoted as $\Pi(\xi,t)$, thus $P(n,t)=\Pi(\xi,t)$. After the change of variables (\ref{ansatz}), the left-hand side of the master equation (\ref{master}) becomes: 
\begin{equation}
\frac{dP}{dt} = \frac{\partial \Pi}{\partial t} - \sqrt{N}\frac{\partial \Pi}
{\partial \xi}\frac{d\phi}{dt}.
\label{LHS}
\end{equation}
The right-hand side of the master equation in the form (\ref{var_master}) can also be written in terms of $\phi$ and $\xi$ by (i) eliminating $n$ in the transition rates (\ref{n_to_n+1}) and (\ref{n_to_n-1}) using Eq.~(\ref{ansatz}), and (ii) noting that the step-operators may be written as
\begin{equation}
{\cal E}^{\pm} = 1 + \sum^{\infty}_{\ell=1} \frac{(\pm1)^{\ell}}{\ell!}
\frac{1}{N^{\ell/2}}\frac{\partial^{\ell}}{\partial \xi^{\ell}}.
\label{Taylor_step}  
\end{equation} 

Equating the left-hand and right-hand sides of the master equation, after rescaling time by introducing $\tau = t/N$, we can match inverse powers of $N^{1/2}$ to obtain a set of equations for the dynamics of the process. This is carried out explicitly in Appendix~\ref{AppA}. At leading order --- obtained by matching the coefficients of $N^{-1/2}$ --- one finds the equation: 
\begin{equation}
\frac{d\phi}{d\tau} = r\phi \left( 1 - \frac{\phi}{K} \right),
\label{meanfield}
\end{equation}
where $r=b-d$ and $K=(b-d)/(b+c)$. This is the logistic equation, which could have been guessed as the deterministic limit of the model, even if the identification of the constant $K$ is not so obvious. This has the solution:
\begin{equation}
\phi(\tau) = \frac{K\phi_0}{\left[K-\phi_0\right]e^{-r\tau} + \phi_0},
\label{soln}
\end{equation}
where $\phi_0 \equiv \phi(0)$. This has the required feature that, as long as $\phi_0 \neq 0$, then $\phi(\tau) \to \phi^{*}$ as $t \to \infty$, where $\phi^{*} = K$ is the non-trivial fixed point.

Once the leading-order contributions have been extracted the left-hand side is simply $\partial \Pi/\partial \tau$, but the right-hand side contains derivatives of $\Pi$ with respect to $\xi$ of all orders. The resulting equation has the general structure:
\begin{eqnarray}
& & \frac{\partial \Pi}{\partial \tau} = \sum^{\infty}_{k=1} \frac{1}{(k+1)!}
\frac{1}{N^{(k-1)/2}} f_{k+1}(\phi) 
\frac{\partial^{k+1} \Pi}{\partial \xi^{k+1}}
\nonumber \\
& & + \sum^{\infty}_{k=1} \frac{1}{k!} \frac{1}{N^{(k-1)/2}} g_{k}(\phi)
\frac{\partial^{k} }{\partial \xi^{k}} \left[ \xi \Pi \right] \nonumber \\
& &  + \sum^{\infty}_{k=2} \frac{1}{(k-1)!} \frac{1}{N^{(k-1)/2}} h_{k-1}
\frac{\partial^{k-1} }{\partial \xi^{k-1}}  \left[ \xi^2 \Pi \right],
\label{nonlin_FPE}
\end{eqnarray}
where the explicit form of the functions $f_{k}, g_{k}$ and the constant $h_{k}$ are given in Appendix~\ref{AppA}.

To proceed in the analysis we introduce the moment of order $q$ of the distribution $\Pi$:
\begin{equation}
\langle \xi^q \rangle = \int \xi^q\,\Pi (\xi)\,d \xi.
\label{moment_def}
\end{equation}
From the generalized Fokker-Planck equation~(\ref{nonlin_FPE}) one can obtain a set of ordinary differential equations for the coupled evolution of the moments of the distribution $\Pi$. The method is straightforward~\cite{Vankampen} and consists in multiplying both sides of Eq.~(\ref{nonlin_FPE}) by $\xi^q$ and integrating by parts over the variable $\xi$. One finds
\begin{eqnarray}
\begin{split}
&\frac{d \langle \xi^q \rangle }{d\tau}=
\sum_{k=1}^{q-1}\frac{(-1)^{k+1}
q! f_{k+1}(\phi) \langle \xi^{q-(k+1)} \rangle }{(k+1)! N^{(k-1)/2} (q-(k+1))!}
\\&+\sum_{k=1}^{q} \frac{(-1)^{k}
q! g_k \langle \xi^{q-k+1} \rangle }{k!N^{(k-1)/2}(q-k)!}
\\&+\sum_{k=2}^{q+1}\frac{(-1)^{k-1}
q! h_{k-1} \langle \xi^{q-(k-1)+2} \rangle }{(k-1)!(q-(k-1))!N^{(k-1)/2}}.
\end{split}
\label{moments}
\end{eqnarray}

It is clear from Eq.~(\ref{moments}), that the equation for $d \langle \xi^q \rangle /d\tau$ depends on $\langle \xi^{q+1} \rangle$ and so the system of equations does not close at any finite value of $q$. The general way to proceed in such cases is to impose some form of truncation which will eventually lead to a closed, self-consistent set of equations. For the case of the model we are investigating here, the fact that Eq.~(\ref{moments}) is an expansion in powers of $1/\sqrt{N}$, suggests the possibility of implementing a rather natural truncation scheme, which we will return to in Section~\ref{simulations}. Once complemented by a particular closure scheme, the system of Eqs.~(\ref{moments}), up to a given value of $q$, can be solved numerically and the estimated moments used to reconstruct the distribution via Fourier inversion, as discussed in Section~\ref{simulations}.

\subsection{The WKB expansion}
\label{WKB_method}
The second method is to use a WKB approximation. This approximation is well-known in the study of ordinary differential equations~\cite{WKB}, but has also been extensively used in the analysis of rare events in stochastic systems, both by mathematicians~\cite{FreidlinWentzell}, and theoretical physicists using path-integrals~\cite{zinn-justin}. Here will look at a variant of the method which starts off from the discrete state master equation, rather than from the Fokker-Planck equation and will follow the methodology described in Refs.~\cite{Bressloff} and \cite{BlackMcKane}. 

The approximation involves a different form of scaling to that described in Section~\ref{vK_method}. Now, the starting point is the master equation (\ref{master}), but written in terms of $x=n/N$ and $N$, rather than $n$. Specifically, we write $P(n,t)=P(Nx,t)=\Pi(x,t)$ and $T_{\pm}(n \pm 1 | n)=N\Omega_{\pm}(x)$, so that:
\begin{equation}
\Omega_{+}(x)=bx(1-x), \ \ \Omega_{-}(x)=x(d+cx).
\label{Omegas_defn}
\end{equation}
We assume that $N$ is sufficiently large that $x$ is effectively continuous. Looking for the quasi-stationary solution of the master equation (\ref{master}), in the basin of the attraction of the stable fixed point, leads to the following equation:
\begin{equation}
\begin{split}
&\Omega_{-}\Big(x+\frac{1}{N}\Big) \Pi\Big(x+\frac{1}{N}\Big) + \Omega_{+}\Big(x-\frac{1}{N}\Big) \Pi\Big(x-\frac{1}{N}\Big)
\\& - \left[ \Omega_{-}(x)+\Omega_{+}(x) \right] \Pi(x) = 0.
\end{split}
\label{ME_Omega}
\end{equation}

To solve the above  time-independent equation for the stationary distribution, we apply the WKB approximation~\cite{Bressloff,BlackMcKane} by assuming the following form for $\Pi(x)$:
\begin{equation}
\Pi(x)=K(x)\exp(-NS(x))\Big[1+O\Big(\frac{1}{N}\Big)\Big],
\label{ansatz_WKB}
\end{equation}  
where both $S(x)$ and $K(x)$ are of the order of unity. Substituting (\ref{ansatz_WKB}) into (\ref{ME_Omega}), Taylor expanding with respect to $N^{-1}$  and collecting together the leading order terms yields
\begin{equation}
\sum_{r= \pm 1}  \Omega_r(x) \left[\exp(rS'(x)-1) \right]=0,
\label{HJ}
\end{equation} 
where $S'(x)=dS(x)/dx$. The above equation can be seen as a stationary Hamilton-Jacobi equation $H(x,S'(x))=0$ for an action $S$ with Hamiltonian 
\begin{equation}
H(x,p)=\sum_{r= \pm 1} \Omega_r(x)[\exp(rp)-1],
\label{hamilton1}
\end{equation} 
where $p=S'(x)$. From Eq.~(\ref{hamilton1}) one gets the following Hamilton's equations:
\begin{eqnarray}
\dot{x} &=& \frac{\partial H}{\partial p} = \sum_{r= \pm 1} r\,\Omega_{r}(x) \exp(rp), \nonumber \\
\dot{p} &=& -\frac{\partial H}{\partial x}=-\sum_{r= \pm 1}[\exp(rp)-1] \frac{\partial \Omega_r(x)}{\partial x},
\label{hamiltonsystem1}
\end{eqnarray}
where the dot denotes differentiation with respect to time.

From the solution of these equations one finds the so-called fluctuation trajectories $x_f$ and the corresponding momenta $p_f$. For the zero-energy solution $H=0$ that we are interested in, the action calculated along a given fluctuation trajectory, starting at the fixed point at time $t_0$, is given by
\begin{equation}
S_f=\int_{t_0}^t p_f \dot{x}_f dt' ,
\label{def_action}
\end{equation}
and so can be calculated from a knowledge of $p_f$ and $\dot{x}_f$.

From Hamilton's equations (\ref{hamiltonsystem1}) one can readily check that a trivial solution, $p_f=0$ exists, provided that
\begin{equation}
\label{det_limit} 
\quad\dot{x}=\Omega_{+}(x)-\Omega_{-}(x),
\end{equation}
where from now on we drop the subscript $f$.

This is customarily called the relaxation trajectory and corresponds to the deterministic ($N \to \infty$) approximation of the model. In fact, multiplying both sides of the original master equation by $n$ and summing over all possible states one finds that
\begin{equation}
\frac{d \langle n \rangle }{dt}=N\Omega_{+}(x)-N\Omega_{-}(x),
\end{equation}
where $\langle n \rangle =\sum_n n P(n,t)$. Dividing then by $N$ and taking the limit $N \to \infty$, one finds Eq.~(\ref{det_limit}). This equation is nothing else but the logistic equation, and the relaxation trajectory eventually converges to the stable fixed point.

In this paper we are interested rather in the solution of the Hamilton's equations (\ref{hamiltonsystem1}) with $p \ne 0$. This enables us to explore trajectories which not allowed in the deterministic limit. This solution, which has a non-trivial value of the momentum, can be found by setting $H=0$ in Eq.~(\ref{hamilton1}), which leads to a quadratic equation in $e^{p}$. Solving this equation yields
\begin{equation}
p=\ln\frac{\Omega_{-}(x)}{\Omega_{+}(x)};\quad \dot{x}=\Omega_{-}(x)-\Omega_{+}(x).
\label{p_x}
\end{equation}

The actions of these paths can now be calculated. In the case $p = 0$ it is simply zero, and in the case given by Eq.~(\ref{p_x}) one finds that
\begin{eqnarray}
S(x) - S(x^*) &=& \frac{(d+cx) \ln(d+cx)}{c}-\frac{(d+cx)}{c} \nonumber \\
&+& \frac{(b-bx)\ln(b-bx)}{b}-\frac{(b-bx)}{b} \nonumber \\
&+& \frac{(d+c)}{c}-\frac{(c+d)}{b} \ln \left[\frac{b(d+c)}{c+b}\right]. 
\nonumber \\
\label{finalexpressionS}
\end{eqnarray}
Here we have used the expression $x^* = (b-d)/(b+c)$ for the non-trivial fixed point, found from the condition $\Omega_{+}(x^*)=\Omega_{-}(x^*)$. 

The next-to-leading terms in the WKB approximation (\ref{ansatz_WKB}) is the prefactor $K(x)$. It is found~\cite{Bressloff,BlackMcKane} to obey the equation
\begin{equation}
\frac{\partial H}{\partial p} \frac{K'}{K} = -\frac{1}{2} p' \frac{\partial^2 H}{\partial p^2}-\frac{\partial^2 H}{\partial p \partial x}.
\label{K}
\end{equation}  
For the case of interest here, $p \ne 0$, with $p(x)$ specified by equation (\ref{p_x}), one can solve equation (\ref{K}) to obtain:
\begin{equation}
K(x)=\frac{A}{\sqrt{\Omega_+(x)\Omega_-(x)}},
\label{eq_for_K}
\end{equation}
where $A$ is a constant to be determined. For the sake of completeness we also give the expression for $K(x)$ which applies to the case $p=0$. It is 
\begin{equation}
K(x)=\frac{B}{\Omega_+(x)-\Omega_-(x)},
\end{equation}
where $B$ is another constant.

We can find the constant $A$ by normalising the probability distribution. If we expand the quasi-stationary distribution $\Pi(x)$ on the trajectory with $p \neq 0$ and in the vicinity of the non-trivial fixed point $x=x^*$ we have that
\begin{equation}
\begin{split}
&\Pi(x)\approx\frac{A}{\sqrt{\Omega_+(x^*)\Omega_-(x^*)}}\exp\Big[-NS(x^*)\\&-\frac{N}{2}S''(x^*)(x-x^*)^2\Big],
\end{split}
\end{equation}
where use had been made of the fact that $S'(x^*)=\ln(\Omega_-(x^*)/\Omega_+(x^*))=0$. A straightforward calculation allows us to write:
$$S''(x^*)=(b+c)^2/b(c+d),$$ 
and therefore:
\begin{equation}
\Pi(x)\approx\frac{A}{\Omega_+(x^*)}\exp \left[-NS(x^*) \right]\,\exp\Big(-\frac{\lambda}{2}(x-x^*)^2\Big),
\end{equation}
where $\lambda=N(b+c)^2/b(c+d)$. By imposing the normalization condition for $\Pi(x)$ we obtain the following expression for $A$: 
\begin{equation}
A=\sqrt{\frac{\lambda}{2\pi}}\exp\left[NS(x^*)\right]\Omega_+(x^*),
\end{equation}
which eventually yields
\begin{equation}
\begin{split}
&\Pi(x)=\sqrt{\frac{NS''(x^*)}{2\pi}}\Big[\frac{\Omega_+(x^*)\Omega_-(x^*)}{\Omega_+(x)\Omega_-(x)}\Big]^{1/2} \\& \times \exp({-N[S(x)-S(x^*)]}).
\end{split}
\label{P_WKB}
\end{equation}

In conclusion, we have derived a closed analytical expression for the quasi-stationary distribution $\Pi$, using the WKB approximation procedure. In the next section we will test the adequacy of formula (\ref{P_WKB}), as well as the corresponding result obtained within the framework of the van Kampen system-size expansion, by performing a direct comparison with the outcome of stochastic simulations.

\section{Numerical simulations}
\label{simulations}
In Section~\ref{model}, by extending the van Kampen system-size expansion beyond the Gaussian order, we obtained a set of coupled differential equations for the moments of the distribution of fluctuations. The knowledge of these moments enables us in principle to reconstruct the corresponding distribution via a standard Fourier inversion. In this Section we will compare this theoretical prediction, along with the WKB result (\ref{P_WKB}), and assess their validity through direct simulation of the stochastic dynamics. 

While it is straightforward to display the result of the WKB analysis, some comments are in order concerning the interpretation of the calculation based on the van Kampen expansion. To find the stationary distribution of fluctuations we consider the ensemble of the first $q$ moments, and impose a truncation in the van Kampen expansion by omitting the term proportional to $\langle \xi^{q+1} \rangle /\sqrt{N}$ in the equation for $d \langle \xi^q \rangle/d \tau$. This is the $k=2$ contribution in the last sum on the right-hand side of equation (\ref{moments}). 

In principle, one cannot formally drop such a term, but we can assess its importance since we are working within a large system size (large $N$) approximation. It is found that the structure of the governing equations means that the effect of neglecting a contribution proportional to $1/\sqrt{N}$ in the equation for the $q$-th moment, becomes rapidly less important as the order of the moments decreases. So by taking $q$ large enough, we would expect that the errors made in the estimates of a significant number of the lowest order moments would be negligible. However, these arguments are heuristic, and ultimately only a comparison with simulations will provide an a posteriori validation of the proposed approximation. 

As an additional remark, we recall that we are interested here in the the asymptotic distribution of fluctuations around the non-trivial fixed point of the deterministic dynamics. This implies setting the derivatives $d \langle \xi^q \rangle / d \tau$ in Eq.~(\ref{moments}) to zero, after having imposed the fixed point condition $\phi=\phi^*$ on the functions $f_k$ and $g_k$, defined in Appendix A. The system of differential equations for the evolution of the moments is hence transformed into an algebraic system that can be readily solved by matrix inversion. The distribution of fluctuations is finally determined by Fourier inverting the corresponding moment-expansion. In the following, we will report results obtained when considering the first thirty moments in the expansion, i.e., taking the first $q=30$ algebraic equations for the stationary moments. 

\begin{figure}[htcp]
\includegraphics[scale=0.4]{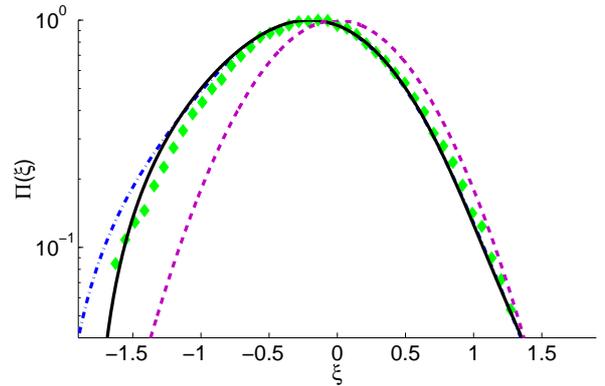} 
\caption{The (green) diamonds represent the distribution in lin-log scale rebuilt from Gillespie's algorithm simulation. The (purple) dashed line, the (blue) dot-dashed line and the (black) solid line show the predicted profile obtained with the first thirty moments and Eq.~(\ref{moments}) truncated at order one, $1/N^{3/2}$ and $1/N^2$, respectively. The parameters used are: $N=1000\quad c=0.5\quad b=0.3\quad d=0.2$. For this choice of the parameters the fixed point of the deterministic dynamics is $\phi^*=0.125$.}
\label{vkapprox}
\end{figure}

In Fig.~\ref{vkapprox} we compare the distributions of fluctuations obtained  via the van Kampen procedure, at different order of approximations of Eq.~(\ref{moments}). The usual van Kampen system-size expansion assumes a Gaussian probability distribution and corresponds to taking only the $N^{0}$ terms in Eq.~(\ref{moments}), that is, only the $k=1$ terms in the first two sums on the right-hand side of this equation. This is displayed using a dashed line (purple online). The dot-dashed (blue) line and the solid (black) line refer respectively to keeping terms up to and including $1/N^{3/2}$ and $1/N^{2}$, respectively, in Eq.~(\ref{moments}). The (green) diamonds represent the distribution rebuilt from direct stochastic simulations, based on the Gillespie's algorithm, for the system. When the order of the approximation is increased, the theoretical distribution tends to adjust well to the numerical profile. 


\begin{figure}[htcp]
\includegraphics[scale=0.4]{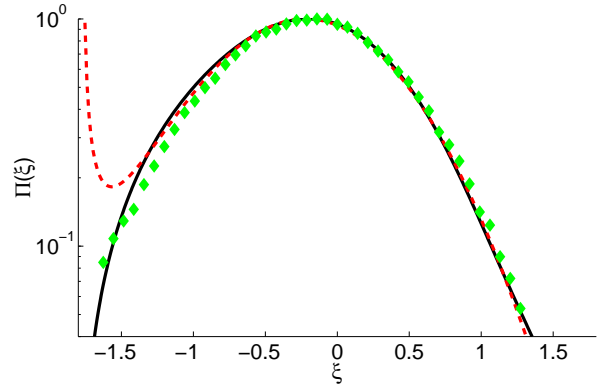} 
\caption{The (green) diamonds represent the distribution in lin-log scale rebuilt from Gillespie's algorithm simulation. The (black) solid line represents the van Kampen approximation with the first thirty moments, and the expansion taken to order of $1/N^{2}$. The (red) dashed line is the WKB approximation. The parameters used are: $N=1000\quad c=0.5\quad b=0.3\quad d=0.2$. For this choice of the parameters the fixed point of the deterministic dynamics is $\phi^*=0.125$.}\label{distribuzione}
\end{figure}

In Fig.~\ref {distribuzione} the performance of the van Kampen system-size expansion and the WKB scheme are compared with each other and with the results of simulations. The agreement between the two schemes and the numerics is satisfactory, with the overall skewness of the distribution appearing to be correctly captured by the approximations. The WKB solution tends to deviate from the expected profile for negative values of the fluctuation $\xi$ (at system-sizes of a few tens of individuals), while the van Kampen approximation, at the order of the expansion that we are working at, still proves to be adequate. The fact that the van Kampen scheme matches the simulated data, constitutes an a posteriori validation of the closure strategy implemented.

 
\section{Conclusion}
\label{conclude}
In this paper we have investigated the distribution of fluctuations in a stochastic model which was chosen to give the logistic equation as its deterministic limit. This equation has the necessary features of a stable non-trivial fixed point and an unstable trivial fixed point required to explore extinction dynamics. Stochastic fluctuations then take the system from the vicinity of the non-trivial fixed point of the deterministic dynamics, to the trivial fixed point, where all the individuals have become extinct. For sufficiently large population sizes, extinctions will be rare and the state of the system will fluctuate about the non-trivial fixed point for long time. 

In this situation the van Kampen system-size expansion constitutes a powerful analytical tool to estimate the distribution of fluctuations. Essentially the method gives the deterministic equations to leading order in an expansion in (inverse) system-size. The linear stochastic corrections found at the next-to-leading order correspond to Gaussian fluctuations, an excellent approximation to the dynamics if extinctions are so rare as to be negligible. When the size of the population is reduced, non-Gaussian traits prove crucial and one needs to go beyond this conventional order of approximation to eventually resolve the skewness of the distribution.  

Alternatively, a WKB-like perturbation scheme can be implemented to derive a closed analytical approximation for the distribution of fluctuations. The method consists of postulating that the dominant contribution to the probability distribution behaves for large system size, $N$, as $e^{-N S(x)}$ where $S(x)$ is a solution of a Hamilton-Jacobi equation found from the carrying out the WKB analysis. The corresponding Hamiltonian can be used to describe the extinction trajectories, and consequently the distribution of stochastic fluctuations. 

The aim of this paper is to explore the connection between the van Kampen system-size expansion at higher order and the WKB method. The two methods have a different basis, and it is therefore interesting to assess their respective predictive ability in regimes where extinctions are important. To this end, we have carried out the explicit calculations involved in these two methods, for the specific case of a stochastic model of the logistic type described above. The theoretical predictions have been compared to the results of the numerical simulations, obtained  by solving the relevant stochastic model via the standard Gillespie algorithm~\cite{gillespie}. In both cases the agreement is satisfactory.

The advantage of the WKB method over the generalized van Kampen expansion is that the former enables one to recover a closed analytical expression for the distribution of fluctuations. In contrast, the latter requires working with a large set of algebraic equations for the moments of the distribution, a step that can be only performed numerically. In addition, the extended van Kampen method must be accompanied by a dedicated truncation strategy, to get a fully consistent set of equations for the unknown moments. The validity of this closure can only be tested a posteriori by a direct comparison with numerical, or experimentally available, data. On the other hand the WKB approximation is unable to capture the behaviour near the extinction boundary; to do this it would need to be matched to a boundary-layer solution calculated outside of the WKB scheme~\cite{Bressloff,BlackMcKane}. However there is no need to do this in the current case, since the van Kampen expansion taken to higher-order fulfills this role. Thus, when used in conjunction with one another, the two approaches show that the extinction dynamics can be correctly captured.

\bigskip

\begin{appendix}

\section{The system-size expansion to higher orders}
\label{AppA}
In this Appendix we give some of the intermediate steps in the derivation of the van Kampen system-size expansion to all orders in $N^{-1/2}$.

The starting point is the substitution of the Taylor series expansion given by Eq.~(\ref{Taylor_step}) into the master equation (\ref{var_master}), while using the van Kampen ansatz Eq.~(\ref{ansatz}). For example, the first term in the master equation gives
\begin{eqnarray}
\begin{split}
&\left( {\cal E}^{+} - 1 \right) [T(n-1|n)P_n(t)]=\Big(\sum_{k=1}^{\infty}\frac{1}{k!}
\frac{1}{N^{k/2}}\frac{\partial^k}{\partial\xi^k}\Big)
\\&\Big[(d\phi+d\xi/\sqrt N+c\phi^2+2c\phi\xi/\sqrt
N+c\xi^2/N)\Pi(\xi)\Big].\nonumber
\end{split}
\label{firstterm}
\end{eqnarray}
It is convenient to group these terms according to the power of $\xi$ which multiplies $\Pi$, as follows:
\begin{eqnarray}
\begin{split}
&\Big(\sum_{k=1}^{\infty}\frac{1}{k!}\frac{1}{N^{k/2}}\frac{\partial^k}{\partial\xi^k}(d\phi+c\phi^2)\Pi\Big)\\&
+\Big(\sum_{k=1}^{\infty}\frac{1}{k!}\frac{1}{N^{k/2+1/2}}\frac{\partial^k}{\partial\xi^k}(2c\phi\xi+d\xi)\Pi\Big)\\&
+\Big(\sum_{k=1}^{\infty}\frac{1}{k!}\frac{1}{N^{k/2+1}}\frac{\partial^k}{\partial\xi^k}\Big[c\xi^2\Big]\Pi\Big).
\end{split}
\end{eqnarray}

The lowest order contribution is the $k=1$ term in the first sum, and this matches the second term in Eq.~(\ref{LHS}) (after the time-rescaling to introduce $\tau$). Omitting this term in the first sum therefore gives the contribution to $\partial \Pi/\partial \tau$ from $({\cal E}^{+} - 1) [T(n-1|n)P_n(t)]$. Carrying out the same steps for the other term in Eq.~(\ref{var_master}), and taking out a factor of $N^{-1}$ from the rescaling, gives the following expression for $\partial \Pi/\partial \tau$:
\begin{eqnarray}
& & \sum_{k=2}^{\infty}\frac{1}{k!}\frac{1}{N^{k/2-1}} \left[ d\phi+c\phi^2 + (-1)^k(b\phi-b\phi^2)\right] \frac{\partial^k \Pi}{\partial\xi^k} + \nonumber \\
& & \sum_{k=1}^{\infty}\frac{1}{k!}\frac{1}{N^{k/2-1/2}} \left[ 2c\phi+d+(-1)^k(b-2b\phi)\right] \frac{\partial^k \left[ \xi \Pi \right]}{\partial\xi^k} 
\nonumber \\
& & +\sum_{k=1}^{\infty}\frac{1}{k!}\frac{1}{N^{k/2}} \left[ c - (-1)^k b \right] 
\frac{\partial^k \left[ \xi^2 \Pi \right]}{\partial\xi^k}.
\end{eqnarray}
If we now introduce the simplifying notation
\begin{eqnarray}
f_k(\phi) &=& \left[ d\phi+c\phi^2+(-1)^k(b\phi-b\phi^2)\right], \nonumber \\
g_k(\phi) &=& \left[ 2c\phi + d + (-1)^k(b - 2b\phi) \right],
\nonumber \\
h_k &=& \left[ c - b(-1)^k \right],
\label{fgh}
\end{eqnarray}
we obtain the generalized Fokker-Planck equation (\ref{nonlin_FPE}) of the main text.

\end{appendix}


\begin{thebibliography}{99}
\bibitem{MckaneEcologyTrends} A. J. Black and A. J. McKane, \textit{Trends in Ecology and Evolution} {\bf 27}, 337 (2012).
\bibitem{McKanePRL} A. J. McKane and T. J. Newman, \textit{Phys. Rev. Lett.} \textbf{94}, 218102 (2005).
\bibitem{dipatti} T. Dauxois, F. Di Patti, D. Fanelli and A.~J. McKane, \textit{Phys. Rev. E} \textbf{79} 036112 (2009).
\bibitem{McKane} C. Lugo and A.~J. McKane, \textit{Phys. Rev. E} \textbf{78}, 051911 (2008).
\bibitem{De_Anna} P. de Anna, F. Di Patti, D. Fanelli, A.~J. McKane and T. Dauxois, \textit{Phys. Rev. E} \textbf {81}, 056110 (2010).
\bibitem{McKaneBiancalaniRoger} A. J. McKane, T. Biancalani and T. Rogers \textit{Bull. Math. Biol.} \textbf{76}, 895 (2014). 
\bibitem{woolley} L. J. Shumacher, T. E. Woolley and R. E. Baker, \textit{Phys. Rev. E} {\bf 87}, 042719 (2013).
\bibitem{Vankampen}N.G. van Kampen. \textit{Stochastic Processes in Physics and
Chemistry} (Elsevier, Amsterdam, 2007). Third edition.
\bibitem{grima} R. Grima, \textit{Phys. Rev. Lett.} \textbf {102}, 218103 (2009).
\bibitem{grima2011} R. Grima, P. Thomas and A.~V. Straube \textit{J. Chem. Phys.} \textbf{135}, 084103 (2011).
\bibitem{cianci_voter} C. Cianci, F. Di Patti and D. Fanelli \textit{Europhys. Lett.} \textbf {96}, 50011 (2011).
\bibitem{cianci_epjst} C. Cianci, F. Di Patti, D. Fanelli and L. Barletti \textit{Eur. Phys. J. Special Topics} \textbf {212}, 5 (2012).
\bibitem{thomas} P. Thomas and R. Grima, \textit{Phys. Rev. E} \textbf {92}, 012120 (2015).
\bibitem{touchette} H. Touchette, \textit{Physics Report} \textbf {478}, 1 (2009).
\bibitem{zinn-justin} J. Zinn-Justin, \textit{Quantum Field Theory and Critical Phenomena} (Clarendon Press, Oxford, 2002). Fourth edition.
\bibitem{FreidlinWentzell} M. I. Freidlin and A. D. Wentzell, \textit{Random Perturbations of Dynamical Systems} (Springer, New York, 1984).
\bibitem{gillespie}D.~T. Gillespie. \textit{J. Comput. Phys.} {\bf 22}, 403 (1976); 
D.~T. Gillespie. \textit{J. Phys. Chem.} {\bf 81}, 2340 (1977).
\bibitem{WKB} C.~M. Bender and S.~A. Orszag, \textit{Advanced Mathematical Methods for Scientists and Engineers} (McGraw-Hill, 1978).
\bibitem{Bressloff} P.C. Bressloff, \textit{Phys. Rev. E} \textbf {82}, 051903 (2010).
\bibitem{BlackMcKane} A. J. Black and A. J. McKane \textit{J. Stat. Mech.} P12006 (2011).

\end{thebibliography}
\end{document}